\documentclass[12pt]{revtex4}

\usepackage{url}
\usepackage{graphicx,color}% Include figure files
\usepackage[psamsfonts]{amssymb}
\usepackage{amsmath}
\usepackage{indentfirst}

\usepackage{booktabs}
\usepackage{tabularx}

\usepackage{color}

\begin{document}
\title{Effective measure of endogeneity for the Autoregressive Conditional Duration point processes via
mapping to the self-excited Hawkes process}

\author{V. Filimonov}
\email{vfilimonov@ethz.ch}
\affiliation{Department of Management, Technology and Economics, ETH Z\"{u}rich, Scheuchzerstrasse 7, CH-8092 Z\"{u}rich, Switzerland}

\author{S. Wheatley}
\email{wspencer@ethz.ch}
\affiliation{Department of Management, Technology and Economics, ETH Z\"{u}rich, Scheuchzerstrasse 7, CH-8092 Z\"{u}rich, Switzerland}

\author{D. Sornette}
\email{dsornette@ethz.ch}
\affiliation{Department of Management, Technology and Economics, ETH Z\"{u}rich, Scheuchzerstrasse 7, CH-8092 Z\"{u}rich, Switzerland}

\date{\today}

\begin{abstract}
In order to disentangle the internal dynamics
from exogenous factors within the Autoregressive Conditional Duration (ACD) model,
we present an effective measure of endogeneity. Inspired from 
the Hawkes model, this measure is defined as the average fraction of 
events that are triggered due to internal feedback mechanisms within the total population. 
We provide a direct comparison of the Hawkes and ACD models based 
on numerical simulations and show that our effective measure
of endogeneity for the ACD can be mapped onto the ``branching ratio''  of the Hawkes model. 
\end{abstract}

\maketitle

%=============================================================================== 
%===============================================================================
\section{Introduction}

An outstanding challenge in socio-economic systems is to disentangle the internal dynamics
from the exogenous influence. It is obvious that any non-trivial system is both subject to external shocks
as well as to internal organizational forces and feedback loops. In absence of external influences,
many natural and social systems would regress or die, however the internal mechanisms are of no less importance and can either stabilize or destabilize the system. These systems are continuously subjected to external shocks, forces, noises and stimulations; they propagate and process these inputs in a self-reflexive way. The stability (or criticality) of these dynamics is characterized by the relative strength of self-reinforcing mechanisms. 

For instance, the brain development and performance is given by both external stimuli and endogenous collective and interactive wiring between neurons. The normal regime of  brain dynamics corresponds to asynchronous firing of neurons with relatively low coupling between individual neurons. However as the coupling strength increases, the internal feedback loops starts playing an increasingly important role in the dynamics, and the system moves towards the tipping point at which abnormal synchronous ``neuronal avalanches'' result in an epileptic seizure~\cite{OxfordTextbookSeizures2012}. As another example,  financial systems are known to be driven by  exogenous idiosyncratic news that are digested by investors and complemented with quasi-rational (sometimes self-referential) behavior. Correlated over-expectations (herding) of investors correspond to the bubble phase that pushes the system towards criticality, where the crash may result as a bifurcation towards a distressed regime ~\cite{Sornette_Crash2003}.

In physical systems at thermodynamic equilibrium, the so-called fluctuation-dis\-sipation theorem relates quantitatively the response of the system to an exogenous (and instantaneous) shock to the correlation structure of the spontaneous endogenous fluctuations~\cite{Stratonovich1992}. In out-of-equilibrium systems, the existence of such relation is still an open question~\cite{Ruelle2004}.
In a given observation set, it seems in general hopeless to separate the contributions resulting from external perturbations
and internal fluctuations and responses. 
However, one would like to understand the interplay between endogeneity and exogeneity (the `endo-exo' problem, for short) 
in order to characterize the reaction of 
a given system to external influences, to quantify its resilience, and explain its dynamics.
Using the class of self-exciting conditional Poisson (Hawkes) processes \cite{Hawkes1971,Hawkes1971_orig}, some progress has recently been made in this direction \cite{Sornette2004_exoendo,DeschatresSornette2005,Sornette2006,SornetteCrane2008}.

In the modeling of complex point processes in natural and socio-economic systems, the Hawkes process \cite{Hawkes1971_orig,Hawkes1971} has become the gold standard due to its simple construction and flexibility. Nowadays, it is being successfully used for modeling sequences of triggered earthquakes~\cite{Ogata1988}; genomic events along DNA~\cite{ReynaudBouret2010}; brain 
seizures~\cite{SorOso10,Lyubushinetal14}; spread of violence~\cite{LewisMohler2010} and crime~\cite{Mohler2011} across some regions; extreme events in financial series~\cite{Embrechts2011} and probabilities of credit defaults~\cite{Errais2010}. In financial applications, the Hawkes processes are most actively used for modeling high frequency fluctuations of financial prices (see for instance~\cite{Bowsher2007,Bauwens2009,FilimonovSornette2012_Reflexivity,Muzy2013hawkes}), however applications to lower frequency data, such as daily, are also possible (see~\ref{apx:financial}).

Being closely related to branching processes~\cite{Harris_Branching2002}, the Hawkes model combines, in a natural and parsimonious way, exogenous influences with self-excited dynamics. It accounts simultaneously for the co-existence and interplay between the exogenous impact on the system and the endogenous mechanism where past events contribute to the probability of occurrence of future events. Moreover, using the mapping of the 
Hawkes process onto a branching structure, it is possible to construct
a representation of the sequence of events according to 
a branching structure, with each event leading to a whole tree of offspring. 

The linear construction of the Hawkes model allows one to separate exogenous events and develop a single parameter, the so-called ``branching ratio'' $\eta$ that directly measures the level of endogeneity in the system. The branching ratio can be interpreted as the fraction of endogenous events within the whole population of events~\cite{Sornette2003geo,FilimonovSornette2012_Reflexivity}. The branching ratio provides a simple and illuminating
characterization of the system, in particular with respect to its fragility and susceptibility to shocks. For $\eta<1$, on average, the proportion $1-\eta$ of events arrive to the system externally, while the proportion $\eta$ of events can be traced back to the influence of past dynamics. As $\eta$ approaches $1$ from below, the system becomes ``critical'', in the sense that its activity is mostly endogenous or self-fulfilling. More precisely, its activity becomes hyperbolically sensitive to external influences. The regime $\eta>1$ corresponds to 
the occurrence of an unbounded explosion of activity nucleated by just a few external events (e.g., news) with non-zero probability.  In any realistic case, when present, this explosion will be 
observable in finite time.  Not only does the Hawkes model provide this valuable parameter $\eta$, but it also amenable to an easy and transparent estimation by maximum likelihood~\cite{Ogata1978,Ozaki1979} without requiring stochastic declustering~\cite{Zhuang2002,Marsan2008}, which is essential in the branching processes' framework but has several limitations~\cite{SornetteUtkin2009}.

However, the Hawkes model is not the only model that describes self-excitation in point processes. In particular, the  Autoregressive Conditional Duration (ACD) model \cite{Engle1997,Engle1998} and the Autoregressive Conditional Intensity (ACI) model \cite{Russell1999} have been introduced and successfully used in econometric applications.  A similar concept was used in the so-called Autoregressive Conditional Hazard (ACH) model~\cite{Hamilton2002}. These processes were designed to mimic properties of the famous Autoregressive Conditional Heteroskedasticity (ARCH) model~\cite{Engle1982} and Generalized Autoregressive Conditional Heteroskedasticity (GARCH) model~\cite{Bollerslev1986} that successfully account for volatility clustering and self-excitation in price time series. Some other modifications of ACD models such as Fractionally Integrated ACD (FIACD)~\cite{Jasiak1998} or Augmented ACD (AACD)~\cite{Fernandes2006} were introduced to account for additional effects (such as long memory) or to increase the flexibility of the model (for a more detailed review, see~\cite{Bauwens2009} and references therein).

In general, all approaches to modeling self-excited point processes can be separated into the classes of Duration-based (represented by the ACD model and its derivations) and Intensity-based approaches (Hawkes, ACH, ACI, and so on), which define a stochastic expression for inter-event durations and intensity respectively. Of all the models, as discussed above, the Hawkes process dominates by far in the class of intensity-based model, and the ACD model -- a direct offspring of the GARCH-family -- is the most used duration-based model. 

Despite belonging to different classes, both models describe the same phenomena and exhibit similar mathematical properties. In this article, we aim to establish a link between the ACD and Hawkes models. We show that, despite the fact that the ACD model cannot be directly mapped onto a branching structure, and thus the branching ratio for this model cannot be derived, it is possible to introduce a parameter 
$\zeta\in[0,1]$ that serves as an effective degree of endogeneity in the ACD model. We show that this parameter shares important properties with the branching ratio $\eta\in[0,1]$ in the framework of the Hawkes model. Namely, both $\zeta$ and $\eta$ characterize stationarity properties of the models, and provide an effective transformation of the exogenous excitation of the system onto its total activity. By numerical simulations, we show that there exists a monotonous relationship between the parameter $\zeta$ of the ACD model and the branching ratio $\eta$ of the corresponding Hawkes model. In particular, the purely exogenous case ($\eta=0$) and 
the critical state ($\eta=1$) are exactly mapped to the corresponding values $\zeta=0$ and $\zeta=1$. We validate our results by goodness-of-fit tests and show that our findings are robust with respect to the specification of the memory kernel of the Hawkes model.

The article is structured as follows. In section~\ref{sec:models}, we introduce the Hawkes and ACD models and briefly discuss their properties. Section~\ref{sec:branching} introduces the branching ratio and relates it to the measure of endogeneity within the framework of the Hawkes model. In section~\ref{sec:endo_ACD}, we discuss similarities between the Hawkes and ACD models, and identify a parameter in the ACD model that can be treated as an effective degree of endogeneity. We support our thesis with extensive numerical simulations and goodness-of-fit tests. In section~\ref{sec:conclusion}, we conclude.

%=============================================================================== 
%===============================================================================
\section{Models of self-excited point processes}\label{sec:models}

Let us define a \emph{univariate point process} of event times $\{t_{i}\}_{i\in\mathbb{N}_{>0}}$ ($t_i>t_j$ for $i>j$) with the \emph{counting process} $\{N(t)\}_{t\geq 0}=\max(i:t_{i}\leq t)$, and the \emph{duration process} of inter-event times $\{\delta t_{i}\}_{i\in\mathbb{N}_{>0}}=t_{i}-t_{i-1}$. Properties of the point process $\{t_{i}\}$ are usually described with the \emph{(unconditional) intensity process} $ \lambda(t)=\lim_{h\downarrow 0}\frac{1}{h}\mathrm{Pr}[N(t+h)-N(t)>0]$ and \emph{conditional intensity process} $ \lambda(t|\mathcal{F}_{t-})=\lim_{h\downarrow 0}\frac{1}{h}\mathrm{Pr}[N(t+h)-N(t)>0|\mathcal{F}_{t-}]$, which is adapted to the natural \emph{filtration} $\mathcal{F}_{t-}=(t_{1},\dots,t_{i}: t<t_i)$ representing the history of the process.

The well-known \emph{Poisson point process} is defined as the point process whose conditional intensity does not depend on the history of the process %$\mathcal{F}_{t-}$ 
and is constant: 
\begin{equation}\label{eq:poisson}
	\lambda(t|\mathcal{F}_{t-})\equiv\lambda(t)=\lambda_0>0,
\end{equation}
The \emph{non-homogenous Poisson process} extends expression \eqref{eq:poisson} to account for time-dependence of both conditional and unconditional intensity functions: $\lambda(t|\mathcal{F}_{t-})\equiv\lambda(t)=\lambda_0(t)>0$. Both homogeneous and non-homogeneous Poisson processes are completely memoryless, which means that the durations $\{\delta t_{i}\}$ are independent from each other and are completely determined by the exogenous parameter (function) $\lambda_0(t)$. 

The \emph{Self-excited Hawkes process} and \emph{Autoregressive Conditional Durations (ACD)} model, which are described in this article, extend the concept of the Poisson point processes by adding path dependence and non-trivial correlation structures. These models represent two different approaches in modelling point processes with memory: the so called \emph{intensity-based} and \emph{duration-based} approaches. As follows from their names, the first approach focuses on models for the conditional intensity function $\lambda(t|\mathcal{F}_{t-})$ and the second considers models of the durations $\{\delta t_{i}\}$. For example, in the context of the intensity-based approach, the Poisson process is defined by equation~\eqref{eq:poisson}.
In the context of the duration-based approach, the Poisson process is defined as the point process whose durations $\{\delta t_{i}\}$ are independent and identically distributed (iid) random variables with exponential probability distribution function $f(\delta t)=\lambda_0\exp(-\lambda_0\delta t)$.
		
%===============================================================================
\subsection{Hawkes Model}
	
The linear Hawkes process~\cite{Hawkes1971_orig,Hawkes1971}, which belongs to the class of intensity-based models, has its conditional intensity $\lambda(t|\mathcal{F}_{t-})$ being a stochastic process of the following general form:
\begin{equation}\label{eq:hawkes_general}
	\lambda(t|\mathcal{F}_{t-})=\mu(t)+\int_{-\infty}^{t}h(t-s)dN(s),
\end{equation}
where $\mu(t)$ is the \emph{background intensity}, which is a deterministic function of time that accounts for the intensity of arrival of \emph{exogenous} events (not dependent on history). A deterministic  \emph{kernel function} $h(t)$, which should satisfy causality ($h(t)=0$ for $t<0$), models the \emph{endogenous} feedback mechanism (memory of the process). Given that each event arrives instantaneously, the differential of the counting process $dN(t)$ can be represented in the form of a sum of delta-functions $dN(t)=\sum_{t_i<t}\delta(t-t_i)dt$, allowing~\eqref{eq:hawkes_general} to be rewritten in the following form:
\begin{equation}\label{eq:hawkes_discr}
	\lambda(t|\mathcal{F}_{t-})=\mu(t)+\sum_{t_i<t} h(t-t_i).
\end{equation}
It can be shown (and we will discuss this point in the following section) that the stationarity of the process~\eqref{eq:hawkes_discr} requires that
\begin{equation}\label{eq:hawkes_stationarity}
	0<\int_0^\infty h(t)dt<1.
\end{equation}

The shape of the kernel function $h(t)$ defines the correlation properties of the process. In particular, the geophysical applications of the Hawkes model, or more precisely of its spatio-temporal extension
called the \emph{Epidemic-Type Aftershock sequence (ETAS)}~\cite{VereJonesOzaki1982,VereJones1970,Ogata1988}, assume in general
a power law time-dependence of the kernel $h(t)$:
\begin{equation}\label{eq:pow}
	h(t)=\frac{K}{(t+c)^\varphi}\chi(t),
\end{equation}
that describes the modified Omori-Utsu law of aftershock rates~\cite{Utsu1961,UtsuOgata1995}. Financial applications~\cite{Hewlett2006,Bowsher2007,Cont2011,FilimonovSornette2012_Reflexivity} traditionally use an exponential kernel
\begin{equation}\label{eq:exp}
	h(t)=a \exp(-t/\tau)\chi(t),
\end{equation}
which has been originally suggested by~\cite{Hawkes1971_orig} and ensures Markovian properties of the model~\cite{Oakes1975}. In both cases, a Heaviside function $\chi(t)$ ensures
the validity of the causality principle. The stationarity condition~\eqref{eq:hawkes_stationarity} requires $Kc^{1-\varphi}/(\varphi-1)<1$ for the power law kernel and $a\tau<1$ for the exponential kernel.

In the present work, we focus on the Hawkes model with an exponential kernel~\eqref{eq:exp} and background intensity $\mu(t)$ that does not depend on time: $\mu(t)\equiv\mu>0$. We introduce a new dimensionless parameter, $\eta=a\tau$, which will be discussed in detail later, which allows us to write the final expression for the conditional intensity
as follows:
 \begin{equation}\label{eq:hawkes_exp}
	\lambda(t|\mathcal{F}_{t-})=\mu+\frac{\eta}{\tau}\sum_{t_i<t}\exp\left(-\frac{t-t_i}{\tau}\right)~.
\end{equation}
Then, the stationarity condition reads $\eta<1$. In order to check the robustness of the results 
presented below, in particular with respect to the choice of the memory kernel, we have also considered a power law kernel~\eqref{eq:pow} with time-independent background intensity $\mu(t)\equiv\mu>0$. Similarly
to the exponential kernel,  the integral from $0$ to $+\infty$ of the memory kernel defines the dimensionless parameter  $\eta=Kc^{1-\varphi}/(\varphi-1)$, which allows us to rewrite the Hawkes model with power law
 kernel as:
 \begin{equation}\label{eq:hawkes_pow}
	\lambda(t|\mathcal{F}_{t-})=\mu+\eta c^{1-\varphi}(\varphi-1)\sum_{t_i<t}\frac{1}{(t-t_i+c)^\varphi},
\end{equation}
Again, the stationarity condition reads $\eta<1$.

%===============================================================================
\subsection{Autoregressive Conditional Durations (ACD) Model}

The class of \emph{Autoregressive Conditional Durations} (ACD) models has been introduced by~\cite{Engle1997,Engle1998} in the field of econometrics to model financial data at the transaction level. The ACD model applies the ideas of the Autoregressive Conditional Heteroskedasticity (ARCH)~\cite{Engle1982} model, which separates the dynamics of a stationary random process into a multiplicative random error term and a dynamical variance that regresses the past values of the process. In the spirit of the ARCH, the ACD model is represented by the duration process $\delta t_{i}$ in the form
\begin{equation}\label{eq:ACD}
	\delta t_{i}= \psi_{i}\epsilon_{i},
\end{equation}
where $\epsilon_{i}$ defines an iid random non-negative variable with unit mean $\mathrm{E}[\epsilon_{i}]=1$, and the function $\psi_{i}\equiv\psi(N(\mathcal{F}_{t-});\theta)$ is the conditional expected duration: $\mathrm{E}[\delta t_{i}|\mathcal{F}_{t-}]= \psi_{i}$. Here, $\theta$ represents the set of parameters of the model. From 
expression~\eqref{eq:ACD}, one can simply derive the conditional intensity of the process~\cite{Engle1998}:
\begin{equation}\label{eq:ACD_intensity}
	\lambda(t|\mathcal{F}_{t-}) = \lambda_{\epsilon}\left( \frac{t-t_{N(t)}}{\psi_{N(t)+1}}\right)\frac{1}{\psi_{N(t)+1}},
\end{equation}
where $\lambda_{\epsilon}(s)$ represents the intensity function of the noise term, $\epsilon_i$. Assuming $\epsilon_i$ to be iid exponentially distributed, one can call this model~\eqref{eq:ACD} an \emph{Exponential ACD} model.

The conditional expected duration $\psi(N(\mathcal{F}_{t-});\theta)$ of the ACD($p$,$q$) model, where $(p,q)$ denotes the order of the model, is defined as an autoregressive function of the past observed durations $\delta t_{i}$ and the conditional durations $\psi_i$ themselves: 
\begin{equation}\label{eq:ACD_Psi}
    \psi_{i}=\omega+\sum_{j=1}^{p}\alpha_{j}\delta t_{i-j}+\sum_{k=1}^{q}\beta_{k}\psi_{i-j},
\end{equation}
where $\omega>0$, $\alpha_{j}\geq0$ and $\beta_{k}\geq0$ are parameters of the model that constitute the set $\theta=\{\omega,\alpha_1,\dots,\alpha_p,\beta_1,\dots,\beta_p\}$. The stationarity condition for the ACD model has the form~\cite{Engle1998}:
\begin{equation}\label{eq:ACD_stationarity}
	\sum_{j=1}^{p}\alpha_{j}+\sum_{k=1}^{q}\beta_{k}<1.
\end{equation}

In the simple ACD(1,1) case that is considered in the present article, equation~\eqref{eq:ACD_Psi} is reduced to:
\begin{equation}\label{eq:ACD_11_Psi}
    \psi_{i}=\omega+\alpha\delta t_{i-1}+\beta\psi_{i-1}.
\end{equation}
Similarly, the conditional intensity~\eqref{eq:ACD_intensity} of the Exponential ACD(1,1) has the form:
\begin{equation}\label{eq:ACD_11_Duration_Process}
    \lambda(t|\mathcal{F}_{t-}) = \frac{1}{\psi_{N(t)+1}} = \frac{1}{ \omega+\alpha \delta t_{N(t)}+\beta \psi_{N(t)} }
\end{equation}
and the stationarity condition~\eqref{eq:ACD_stationarity} reduces to $\alpha+\beta<1$.

%=============================================================================== 
%===============================================================================
\section{The Branching ratio as a measure of endogeneity in the Hawkes model}\label{sec:branching}

The linear structure of the Hawkes process~\eqref{eq:hawkes_discr} with identical functional form of summands $h(t-t_i)$, that depend only on arrival time of a single event $t_i$,
allows one to consider it as a cluster process in which the random process of cluster centers $\{t^{(c)}_i\}_{i\in\mathbb{N}_{>0}}$ is the Poisson process with rate $\mu(t)$. All clusters associated with centers $\{t^{(c)}_i\}$ are mutually independent by construction and can be considered as a \emph{generalized branching process}~\cite{Hawkes1974}, illustrated in figure~\ref{fig:branching}.

\bigskip

[Insert Figure~\ref{fig:branching} here] 

\bigskip

In this context, each event $t_i$ can be either an \emph{immigrant} or a \emph{descendant}. The rate of immigration is determined by the background intensity $\mu(t)$ and results in an exogenous random process. Once an immigrant event occurs, it generates a whole cluster of events. Namely, a zeroth-order event (which we will call the \emph{mother event}) can trigger one or more first-order events (\emph{daughter events}). Each of these daughters, in turn, may trigger several second-order events (the grand-daughters
of the initial mother), and so on. All first-, second- and higher-order events form a cluster and are called descendants (or \emph{aftershocks}) and represent endogenously driven events that appear due to internal feedback mechanisms in the system. 
It should be noted that this mapping of the Hawkes process~\eqref{eq:hawkes_discr} onto the branching structure (figure~\ref{fig:branching}) is possible due to the linearity of the model, and is not valid for nonlinear self-excited point processes, such as the class of nonlinear mutually excited point processes~\cite{BremaudMassoulie1996}, of which the Multifractal stress activation model~\cite{SornetteOuillon2005PRL}
is a particular implementation.

The crucial parameter of the branching process is the \emph{branching ratio} ($n$), which is defined as the average number of daughter events per mother event. Depending on the branching ratio, there are three regimes: (i) \emph{sub-critical} ($n<1$), (ii) \emph{critical} ($n=1$) and (iii) \emph{super-critical} or explosive ($n>1$). Starting from a single mother event (or immigrant) at time $t_1$, the process dies out with probability $1$ in the sub-critical and critical regimes and has a finite probability to explode to an infinite number of events in the super-critical regime. The critical regime for $n=1$ separates the two main regimes and is characterized by power law statistics of the number of events and in the number of generations before extinction~\cite{SaichevSornette2005}. For $n\leq 1$, the process is stationary in the presence of a Poissonian or more generally stationary flux of immigrants.

Being the parameter that describes the clustering structure of the branching process, the branching ratio $n$ defines the relative proportion of exogenous events (immigrants) and endogenous events (descendants or aftershocks). Moreover, in the sub-critical regime, in the case of a constant background intensity ($\mu(t)=\mu=\mbox{const}$), the branching ratio is exactly equal to the fraction of the average number of descendants in the whole population~\cite{Sornette2003geo,FilimonovSornette2012_Reflexivity}. In other words, the branching ratio is equal to the proportion of the average number of endogenously generated events among all events and can be considered as an effective measure of endogeneity of the system.

To see this, let us count separately the rates of exogenous and endogenous events. The rate of exogenous immigrants (zeroth-order events) is equal to the background activity rate: $R_{exo}=\mu$. Each immigrant independently gives birth, on average, to $n$ daughters and thus the rate of first-order events is equal to $r_1=\mu n$. In turn, each first-order event produces, on average, $n$ second-order events, whose rate is equal to $r_2=n r_1=\mu n^2$. Continuing this process ad infinitum and summing over all generations, we obtain the rate of all endogenous descendants:
\begin{equation}\label{eq:R_endo}
	R_{endo}=\sum_{i=1}^{\infty}r_i=\mu\sum_{i=1}^\infty n^i=\frac{\mu n}{1-n},
\end{equation} 
which is finite for $n<1$.  The global rate is the sum of the rates of immigrants and descendants and equal to 
\begin{equation}\label{eq:R_all}
	R=R_{exo}+R_{endo}=\mu+\frac{\mu n}{1-n}=\frac{\mu}{1-n}.
\end{equation}
And the proportion of descendants (endogenously driven events) in the whole system is equal to the branching ratio:
\begin{equation}\label{eq:R_n}
	\frac{R_{endo}}{R}=n.
\end{equation}
Calibrating $n$ on the data therefore provides a direct quantitative estimate of the degree of endogeneity. 

In the framework of the Hawkes model~\eqref{eq:hawkes_discr} with $\mu(t)=\mu=\mbox{const}$, the branching ratio $n$ is easily defined via the kernel $h(t)$:
\begin{equation}\label{eq:n}
	n=\int_0^\infty h(t)dt.
\end{equation}
For the exponential parametrization~\eqref{eq:exp}, the branching ratio, $n=a\tau$, is equal to a 
dimensionless parameter $n\equiv\eta$ previously introduced. The Hawkes framework provides a convenient way of estimating the branching ratio, $n\equiv\eta$, from the observations $\{t_i\}$, using the Maximum Likelihood method, which benefits from the fact that the log-likelihood function is known for Hawkes processes~\cite{Ogata1978,Ozaki1979}. The calibration of the model and estimation of the branching ratio $n$ can then be performed with the numerical maximization of Log-Likelihood function in the parameter space $\{\mu,n,\tau\}$ for the exponential kernel~\eqref{eq:exp} and $\{\mu,n,c,\varphi\}$ for the power law model~\eqref{eq:pow}. Despite being a relatively straightforward calibration procedure, special care should be taken with respect to data processing, choice of the kernel, robustness of numerical methods and stationarity tests as discussed in details in~\cite{FilimonovSornette2013_apparent}.

%=============================================================================== 
%===============================================================================
\section{The Effective degree of endogeneity in the Autoregressive Conditional Durations (ACD) model}\label{sec:endo_ACD}

\subsection{Formal similarities between the ACD and Hawkes models}

Note that the ACD($p$,$q$) and Hawkes models 
operate on different variables with inverse dimensions:
duration $\delta t$ for the ACD($p$,$q$) model and conditional intensity 
$\lambda(t|\mathcal{F}_{t-})$ for the Hawkes model, which is of the order of the inverse $1/\delta t$
of the duration $\delta t$. As a consequence, equations~\eqref{eq:ACD_mean} and~\eqref{eq:R_all} 
apply to different statistics (average durations $\mathrm{E}[\delta t]$ and 
average rate $R=\mathrm{E}[1/\delta t]$).
Moreover, the ACD model cannot be exactly mapped onto a branching structure whereas the Hawkes process can. 
%\todo{As discussed in Sec.~\ref{sec:branching}, the exact mapping of the Hawkes process onto the branching structure results from the additivity of underlying Poisson process and additive form of conditional intensity~\eqref{eq:hawkes_general}. This allows for the intensity to be decomposed into immigration and offspring processes. However, the ACD process has a multiplicative conditional intensity~\eqref{eq:ACD_11_Duration_Process} which cannot be expressed in linear form with a finite number of terms. Therefore it cannot be decomposed into immigration and offspring processes.}

Indeed, the branching structure requires that the conditional probability for an event to occur within the infinitely small interval $[t,t+dt)$ (which is the conditional intensity) should be decomposed into a sum of (1) a (deterministic or stochastic) function of time that represents the immigration intensity and (2) the contributions $f_i$ from each past event $t_i$ that satisfy the following conditions: (i) these contributions should depend only on $t_i$ and be independent from all other events $t_j<t$; (ii) these contributions should exhibit identical 
structure for all events; and (iii) they should satisfy the causality principle. 
Thus, in its general form, a conditional Poisson process 
that can be mapped on (multiple) branching structures if it is described by the following conditional intensity:
\begin{equation}\label{eq:branching_1}
	\lambda(t|\mathcal{F}_{t-})=
	\mu(t)+\sum_{t_i<t} f(t,t_i)\chi(t-t_i),
\end{equation}
where $f(t, t_i)$ is some deterministic function, and $\chi(t)$ is a unit step (Heaviside function). In the context of autoregressive models (such as ACD), the expected waiting time at a given time $t$ is defined as a regressive sum of past durations, which means that the contribution of each event $t_i<t$ to the intensity at time $t$ 
depends on all the events $t_j$ that happened after it ($t_i<t_j<t$). This violates the first principle 
for a branching processes of the independence of distinct branches. For the ACD model, the analysis is also complicated by the structure of the conditional intensity function~\eqref{eq:ACD_intensity}, where past history is influencing the intensity both in a multiplicative way and with a shift in the baseline intensity $\lambda_\epsilon$. One should note that autoregressive intensity models (such as ACI) in general also do not have a branching structure representation due to the problem discussed above.

Despite the differences in their definition and the impossibility of developing an exact mapping onto a branching structure, the ACD model shares many similarities with the Hawkes model
and their point processes exhibit similar degrees of clustering. In particular,
for the ACD defined by expression~\eqref{eq:ACD_Psi}, the combined parameter, 
\begin{equation}\label{eq:zeta}
	\zeta=\sum_{j=1}^p\alpha_j+\sum_{k=1}^q\beta_k
\end{equation}
plays a similar role to the parameter $\eta$ of the Hawkes process with an exponential kernel~\eqref{eq:hawkes_exp}. The similarities start with the stationarity conditions~\eqref{eq:hawkes_stationarity} and~\eqref{eq:ACD_stationarity}, which require $\eta<1$ for the Hawkes model and $\zeta<1$ for the ACD, but go much deeper than the simple idea of ``effective distance'' to a non-stationary regime. 

As we have seen in the previous section, $\eta$ defines the effective degree of endogeneity~\eqref{eq:R_n} that translates the exogenous rate $R_{exo}=\mu$ into the total rate $R_{total}=R_{exo}/(1-\eta)$. 
Similarly, let us study the role of endogenous feedback in the ACD model. 
For $\alpha_j=\beta_k=\zeta=0$, the ACD(0,0) model~\eqref{eq:ACD},\eqref{eq:ACD_Psi} reduces to a simple Poisson process with durations $\delta t_i=\omega\epsilon_i$ having an average value of $\mathrm{E}[\delta t_i]=\omega$, which can be considered as the exogenous factor. When $\alpha_j>0$ and $\beta_k>0$,
there is an amplification of the average durations. Considering the average of eq.~\eqref{eq:ACD_Psi} in the stationary regime ($\mathrm{E}[\delta t_{i-1}]=\mathrm{E}[\delta t_{i}]$ and $\mathrm{E}[\psi_{i-1}]=\mathrm{E}[\psi_{i}]$), and taking into account eq.~\eqref{eq:ACD}, we obtain the following expression for the mean duration in the stationary regime:
\begin{equation}\label{eq:ACD_mean}
	\mathrm{E}[\delta t]=	\frac{\omega}{1-\sum_{j=1}^p\alpha_j-\sum_{k=1}^q\beta_k}\equiv\frac{\omega}{1-\zeta}.
\end{equation}
Equations~\eqref{eq:ACD_mean} and~\eqref{eq:R_all} share the same functional dependence,
with a divergence when the corresponding control parameters $\eta$ and $\zeta$ approach $1$.

\subsection{Empirical dependence of the effective branching ratio ${\hat \eta}$ as a function of $\zeta=\alpha+\beta$ for the ACD($1$,$1$) process}

In order to quantify the similarities between the ACD and Hawkes models
outlined in the previous section, we have performed the following numerical study. 
We simulated realizations of the ACD(1,1) process and calibrated the Hawkes model to it. 
The traditional way of fitting the Hawkes model uses the maximum likelihood method~\cite{Ozaki1979}, 
which is asymptotically normal and asymptotically efficient~\cite{Ogata1978}. 
We have used the R package ``PtProcess''~\cite{Harte2010_PtProcess}, which 
provides a convenient framework for Hawkes models~\eqref{eq:hawkes_discr} 
with arbitrary kernel $h(t)$ and background intensity $\mu(t)$. Then, 
we maximized the likelihood function using a Newton-type non-linear maximization~\cite{Schnabel1986,DennisSchnabel_1987Optimization}. The~\ref{apx:bias}
reports a study of the finite sample bias and efficiency of the Hawkes maximum likelihood estimator.
We find that the
estimation error $|\hat\eta-\eta|$ of the branching ratio (without model error)
measured with the 90\% quantile ranges does not exceed $0.1$ for all values $\eta \leq 0.9$.

More precisely, we want to quantify similarities between control parameters $\zeta$ of the model ACD($p$,$q$) and $\eta$ of the exponential Hawkes model. For this, we have simulated realizations of the ACD process and estimated the parameter $\eta$ from these realizations. The parameter $\omega$ of the ACD($p$,$q$)~\eqref{eq:ACD_Psi} model defines the time scale. Without loss of generality, we let $\omega=1$, which accounts for a linear transformation of time $\tilde t_i=t_i/\omega$ in equations~\eqref{eq:ACD} and~\eqref{eq:ACD_Psi}. 
For the sake of simplicity, 
we present our results for the ACD(1,1) model, for which 
the dimensionless parameter $\zeta$ reduces to $\zeta=\alpha+\beta$. However, our findings are robust to the choice of the order of the ACD model and can be easily generalized to the case of $p,q>1$.
The parameters $\alpha$ and $\beta$ were chosen so that  $\zeta=\alpha+\beta$ spanned $[0,1]$ at 40 equidistant points. For each of the 40 values of $\zeta$, we have generated 100 realizations of the corresponding exponential ACD(1,1) process.  Each realization of 3500 events was generated by a recursive algorithm using eq.~\eqref{eq:ACD_11_Psi}. In order to minimize the impact of edge effects that can bias the estimation of the branching ratio~\cite{FilimonovSornette2013_apparent}, the first 500 points of each realization were discarded. Then, the Hawkes model~\eqref{eq:hawkes_exp} was calibrated on these synthetic datasets.  

For each calibration, we have performed a goodness-of-fit test based on residual analysis~\cite{Ogata1988}, which consists of studying the so-called residual process defined as the nonparametric transformation of the initial time-series $t_i$ into 
\begin{equation}\label{eq:residuals}
	\xi_i=\int_0^{t_i}\hat\lambda(t|\mathcal{F}_{t-})dt,
\end{equation}
where $\hat\lambda(t|\mathcal{F}_{t-})$ is the conditional intensity of the Hawkes process \eqref{eq:hawkes_exp} estimated with the maximum likelihood method. Under the null hypothesis that the data has been generated by the Hawkes process~\eqref{eq:hawkes_exp},  the residual process $\xi_i$ should be Poisson with unit intensity~\cite{Papangelou1972}. Visual analysis involves studying the cusum plot or Q-Q plot and may be complemented with rigorous statistical tests. Under the null hypothesis (Poisson statistics of the residual process $\xi_i$), the inter-event times in the residual process, $\delta\xi_i=\xi_i-\xi_{i-1}$, should be exponentially distributed with CDF $F(\delta\xi)=1-\exp(-\delta\xi)$. Thus, the random variables $U_i\equiv F(\delta\xi_i)=1-\exp(-\delta\xi_i)$ should be uniformly distributed in $[0,1]$. We have performed rigorous Kolmogorov-Smirnov tests for uniformity and provided the corresponding p-values.

We start with a visual comparison of realizations generated with the two models. Figure~\ref{fig:ACDHawkesSeries} presents a comparison of the conditional intensities and durations for (i) simulations of the ACD(1,1) process and, (ii) simulations of the Hawkes process with parameters calibrated to the corresponding ACD process realization. Visual similarities are striking for all four ACD-Hawkes pairs: Total, average, and maximum durations are similar. Moreover, bursts of short and long durations are of similar length. The conditional intensities  fluctuate in a similar range and show qualitatively similar clustering of events, although the ACD conditional intensity is constant between events while the Hawkes decays exponentially. Quantitatively, the distributions of durations also show a large degree of similarity.

\bigskip

[Insert Figure~\ref{fig:ACDHawkesSeries} here] 

\bigskip

Figure~\ref{fig:ACDHawkesSeries} also reveals one important property of the ACD model. Despite the fact that many statistical properties (such as average durations~\eqref{eq:ACD_mean}) are defined by the control parameter $\zeta=\alpha+\beta$, $\alpha$ and $\beta$ have different
impacts on the effective degree of endogeneity $\eta$. For instance, case (B) $\alpha=0.38, \beta=0.13$, and 
case(C) $\alpha=0.13, \beta=0.38$ both have the same $\zeta=0.51$ but 
$\hat\eta=0.52$ for (B) and $\hat\eta=0.22$ for (C). The smaller endogeneity found
in case (C) is compensated by a higher rate of exogenous events ($\hat\mu=0.38$ for (C)
compared with $\hat\mu=0.24$ for (B)), resulting in a ``flatter'' conditional intensity for (C). 

In order to explore this effect  in simulations of the ACD model, for each value of $\zeta=\alpha+\beta$, we considered different relations between $\alpha$ and $\beta$: (i) $\alpha=\beta~ (=\zeta/2)$, (ii) $\beta=0~ (\alpha=\zeta)$, (iii) $\alpha=0~ (\beta=\zeta)$, (iv) $\alpha=3\beta ~(=3\zeta/4)$ and (v) $\beta=3\alpha~ (=3\zeta/4)$. Figure~\ref{fig:HawkesFit} presents the results of the fitting of the Hawkes model on realizations of the ACD(1,1) model in these five cases. The first striking observation 
is the existence of two fundamentally different behaviors observed
for  $\alpha=0$ (case (iii)) versus $\alpha>0$ (cases (i),(ii),(iv),(v)).
For $\alpha=0$, the estimated effective branching ratio $\hat\eta$ is 
$0$ for all values of the control parameter $\zeta=\beta$, as shown
in Figure \ref{fig:HawkesFit}(B). This diagnoses a
completely exogenous dynamics of the ACD process, which is indeed the
expected diagnostic given that, for $\alpha=0$,
eq.~\eqref{eq:ACD} and~\eqref{eq:ACD_11_Psi} reduce to 
\begin{equation}\label{eq:alpha_0}
	\delta t_i=\psi_i\epsilon_i,\quad
	\psi_i=\omega+\beta\psi_{i-1},
\end{equation}
for which the dynamics of the conditional durations $\{\psi_i\}$
is purely deterministic and independent of the realized durations $\delta t_i$,
while the later are entirely driven by the random term $\epsilon_i$.

\bigskip

[Insert Figure~\ref{fig:HawkesFit} here] 

\bigskip

For $\alpha>0$, we find similar non-trivial results.  Figure~\ref{fig:HawkesFit}(B)
shows the effective branching ratio $\hat\eta$ as
a monotonously increasing function of $\zeta$ for all combinations of $\alpha \neq 0$ and $\beta$. 
In cases (ii) ($\beta=0$) and (iv) ($\alpha=3\beta$), the dependence of $\hat\eta$ on $\zeta$ is almost linear for $\zeta<0.5$ and $\zeta<0.9$ respectively and, for higher values of $\zeta$, the convexity increases. In case (i)
($\alpha=\beta$), $\hat\eta$ depends linearly on $\zeta$ for $\zeta>0.4$
with a very good approximation. Finally, in case (v) ($\alpha=3\beta$), 
the curvature of $\hat\eta(\zeta)$ is significant over the range of $0.3<\zeta<0.9$. 
Remarkably, all four dependencies converge to the same value $\hat\eta\approx0.9$ for $\zeta=1$. 

Figure~\ref{fig:HawkesFitUnbias} presents the dependence of the effective branching ratio $\hat\eta$ on the control parameter $\zeta=\alpha+\beta$ after correction of the bias in estimation due to 
finite size effects presented in the~\ref{apx:bias} and summarized in 
figure~\ref{fig:Bias}).  All dependencies of $\hat\eta$ as
a function of $\zeta$ converge to the critical value $\zeta=1$. 

\bigskip

[Insert Figure~\ref{fig:HawkesFitUnbias} here] 

\bigskip

Figure~\ref{fig:contour} generalizes figure~\ref{fig:HawkesFit}(B) by
presenting the dependence of the effective branching ratio $\hat\eta$ (corrected for 
the finite sample bias determined in the~\ref{apx:bias}) on the parameters $\alpha$ and $\beta$ separately.  
As expected, the impact of a change of $\alpha$ is much larger than that of $\beta$.
There is a region, delineated by the dashed line, within which the Hawkes model
is rejected at the 5\% level for the Kolmogorov-Smirnov test.  
For most combinations of $\alpha$ and $\beta$ such that $0.6\lesssim \alpha+\beta \lesssim 0.95$,
the Hawkes model is rejected.  Interestingly, the Hawkes model is not rejected in the case where $\beta$ is kept significantly larger than $\alpha$, and it is only rejected in a small interval in the extreme opposite case where $\beta\equiv0$. The model is often not rejected for large values of $\hat{\eta}$.

\subsection{Differences between the ACD and Hawkes models}

Despite similarities, the Hawkes and ACD models exhibit some important differences. 
Figure~\ref{fig:HawkesFit}A shows that the effective background rate $\hat\mu$
estimated by the Hawkes model is a decreasing function of 
the control parameter $\zeta$. This is an indirect consequence of the 
dependence of the expected duration on $\zeta$
given by expression \eqref{eq:ACD_mean}. 
In contrast to the Hawkes model~\eqref{eq:hawkes_general},  for which the background 
rate $\mu(t)$ completely describes the exogenous impact on the system, 
the parameter $\omega$ of the ACD model~\eqref{eq:ACD_Psi} is not the only factor 
embodying the exogenous activity and there is no strict decoupling
between the exogenous driver $\omega$ and endogenous level $\zeta$ as 
occurs for the parameters $\mu$ and $\eta$ of the Hawkes model. 
In other words, in contrast to the Hawkes model, 
the ACD in its classical form~\eqref{eq:ACD_Psi} does not provide
a clean distinction between exogenous and endogenous activities.
 
Another difference between the Hawkes and ACD models can be observed
in figure~\ref{fig:HawkesFit}(D), which presents a residual analysis of 
the calibration of realizations of the ACD process by the Hawkes model 
using the Kolmogorov-Smirnov test. The null hypothesis that 
the realizations of the ACD process are generated by the Hawkes model 
is rejected at the 5\% confidence level for $\zeta>0.6$ in case (i) ($\alpha=\beta$). 
For case (ii) ($\beta=0$) and (iv) ($\alpha=3\beta$), the null hypothesis
is rejected for even lower $\zeta>0.4$. However, for case (v) ($\beta=3\alpha$),
the null cannot be rejected for almost all values of the control parameter 
$\zeta$, except for a small interval around $\zeta\approx0.8$.

\subsection{Influence of the memory kernel $h(t)$ of the calibrating Hawkes process}

Finally, we need to discuss the choice of the kernel $h(t)$ in
the specification of the Hawkes model~\eqref{eq:hawkes_discr} used
in the calibration of the realizations generated with the ACG process. 
The use of the exponential kernel~\eqref{eq:exp} is a priori justified
by the short memory of the ACD($1$,$1$) process. Indeed, the autocorrelation 
function of the ACD($1$,$1$) model decays exponentially~\cite{Bauwens2009}, and the same 
can be shown explicitly for the GARCH($1$,$1$) model~\cite{Francq2010_GARCH}. 
The choice of a short-memory exponential kernel for the Hawkes model
ensures Markovian properties with a fast decaying autocorrelation function of 
the durations~\cite{Oakes1975}. High p-values of the goodness-of-fit tests for 
parameters $\zeta<0.5$ (see figure~\ref{fig:HawkesFit}) confirm the good mapping between the 
exponential ACD(1,1) and Hawkes processes with an exponential kernel. 

In order to further validate the selection of the exponential kernel of the Hawkes
process, we have compared the calibrations of realizations generated with the ACG process
with the Hawkes model with the exponential kernel~\eqref{eq:exp} and with 
the power law kernel~\eqref{eq:pow}. Since these models have a different number of parameters ($k=3$ and $k=4$ respectively), we compare them using the Akaike information criterion (AIC)~\cite{Akaike1974}.  
The AIC is by far the most popular model comparison criterion used in the point process literature~\cite{Guttorp2012}.  The AIC penalizes complex models by discounting the likelihood function $L$ by the number
$k$ of parameters of the model. Specifically, the AIC suggests selecting the model with a minimum $AIC$ value, where $AIC=2k-2\log L$.

\bigskip

[Insert Table~\ref{tb:exp_pow} here] 

\bigskip

Table~\ref{tb:exp_pow} gives the results for the realizations presented in figure~\ref{fig:ACDHawkesSeries}. 
In terms of likelihood, the exponential and power law kernels give practically 
identical values ($\log L_{exp}\approx\log L_{pow}$). Penalizing model complexity with the AIC widens the gap, and the exponential kernel with one fewer parameters is selected under the AIC. 

Notwithstanding their apparent strong difference,  the estimated background intensities ($\hat\mu$) and branching ratios ($\hat n$) are almost the same for both memory kernels.
This can explained by the fact that the parameters $\hat \varphi$ and $\hat c$
estimated for the power law kernel (\ref{eq:pow})
are such that the later remains very close to an exponential kernel over a large time interval,
as illustrated by figure~\ref{fig:kernels} for case C ($\alpha=0.13$, $\beta=0.38$),
which presents a direct comparison between the exponential 
kernel $\tilde h(t)=h(t)/\eta=\tau^{-1}\exp(-t/\tau)$) and the power law kernel
$\tilde h(t)=c^{1-\varphi}(\varphi-1)(t+c)^{-\varphi}$. The corresponding ML estimates
of their parameters are respectively $\hat\tau=7.76$, $\hat\varphi=105.17$ and $\hat c=816.41$. 
The large value of the estimated exponent $\hat \varphi$ (of the order of 100) implies a fast decay, similar
to an exponential function. Correlatively, the large value of the constant $\hat c$
implies the absence of the hyperbolic range (or ``long tail'') as well.
The almost perfect coincidence is observed for up to times $t \simeq 30$,
over which the kernels $\tilde h(t)$ decay by a factor of almost 50. 
For $t>50$ for which the relative difference between
the  two kernels exceed 20\%, the absolute values of $h(t)$ 
is less than $2\cdot10^{-4}$ so that the contribution of time scales
beyond $t=50$ to the total intensity~\eqref{eq:hawkes_exp},\eqref{eq:hawkes_pow} 
becomes insignificant.

\bigskip

[Insert Figure~\ref{fig:kernels} here] 

\bigskip

%==============================================================================
%==============================================================================
\section{Conclusion}\label{sec:conclusion}

The present article positions itself within the neoclassical financial literature 
that investigates the nature of the mechanisms that drive financial prices.
The benchmark,  called the ``Efficient Market Hypothesis'' (EMH)~\cite{Fama1991,Fama1970,Samuelson1965,Samuelson1973},
holds that markets only reacts to external inputs (information flow) and 
almost instantaneously reflect these inputs in the price dynamics. 
This purely exogenous view on price formation has been 
contradicted by many empirical observations (see for instance 
the original works~\cite{Shiller1981,Cutler1987} and more recent~\cite{Fair2002,SonetteMalevergneMuzy2003,Bouchaud2008}),
which show that only a minor fraction of price movements can be explained by relevant news releases.
This implies a significant role for internal feedback mechanisms.
Using the framework of Hawkes processes, two of us \cite{FilimonovSornette2012_Reflexivity,FilimonovUNCTAD_JIMF2013}
have used the corresponding branching ratio to provide what is, to the best of our knowledge, 
the first quantitative estimate of the degree of endogeneity in financial markets.  
This degree of endogeneity is measured as the proportion of price moves 
resulting from endogenous interactions among the total number of all price moves 
(including both endogenous interactions and exogenous news). These works provided a solid counter-example of short-term ``inefficiency'' of financial markets, which was complemented with the similar confirmation from longer time scales~\cite{HardimanBouchaud2013}. The later work, though, is subjected to a number of numerical biases, as shown in~\cite{FilimonovSornette2013_apparent}, and triggered an ongoing discussion about the nature of long-range memory and criticality.

In this context, the present article expands the quantification of endogeneity
to the class of Autoregressive Conditional Duration (ACD) point processes.
This is done by the introduction of the  
composite parameter $\zeta$ (\ref{eq:zeta}) associated with the parameters
$\alpha_j$ and $\beta_k$, which
control the dependence of the conditional expectated duration
between events as a function of past realized duration and 
past conditional expected duration. We have shown that the parameter $\zeta$
can be mapped onto the branching ratio $\eta$ that directly measures the level of 
endogeneity within the framework of the Hawkes self-excited
conditional Poisson model. This result leads to a novel interpretation
of the various studies that analyzed high-frequency 
financial data with the ACD model. 

An important conclusion derives from our mapping
of the ACD onto the Hawkes process.
Both original works~\cite{Russell1999,Engle1998,Engle2000}
as well as more recent studies reviewed in Refs.~\cite{Pacurar2008,Engle2009} 
have reported estimated parameters $\alpha_j$ and $\beta_k$
that combine to extremely large values of $\zeta$,
often larger than $0.5$, and up to $0.95$. From the 
perspective offered by the present work and in particular
from the mapping of $\zeta$ onto $\eta$,
these empirical findings provide strong support to the hypothesis of 
a dominant endogenous  or ``reflexive''~\cite{Soros_Alchemy1988} 
component in the dynamics of financial markets.

The present work offers itself to a natural extension beyond point processes to the 
class of discrete time processes. There are several successful 
models of self-excitation within a discrete time framework, such as 
AR (auto-regressive), ARMA (auto-regressive moving average)~\cite{Hamilton1994} and 
GARCH models~\cite{Bollerslev1986} and their siblings, as well as the recently
introduced Self-Excited Multifractal (SEMF) model~\cite{FilimonovSornette2011_SEMF}, that extends Quasi-Multifractal models~\cite{SaichevSornette2006,SaichevFilimonov2008Letters} by introducing explicit feedback mechanism.
However, until now, there has been no framework that provides
a direct quantification and estimation of the degree of endogeneity 
present in a given time series. As discussed above, the ACD($p$,$q$) model in fact belongs
to the class of GARCH($p$,$q$) models, though not with normally distributed innovations 
but instead with iid distributed innovations with a Poisson distribution. 
By extension, this suggests a direct application of our present findings 
to GARCH models. This correspondence
will benefit from the elaborate toolbox of calibration methods 
and the detailed accumulated knowledge of the statistical properties
of GARCH models~\cite{HandbookFinancialTimeSeries2009,Francq2010_GARCH}.

\pagebreak

%============================================================================== 
%==============================================================================
\appendix
\section{Financial applications of the Hawkes and ACD models}\label{apx:financial}

Both Hawkes and ACD-type models belong to a class of point processes and describe stochastic arrival times of events of some kind. Since the key variable of these models is the arrival time, selection of what defines an event is extremely important both for numerical analysis and for the diagnostic of the exogenous and endogenous mechanisms. In~\cite{FilimonovUNCTAD_JIMF2013}, a number of endogenous mechanisms that exist in financial markets are listed --- ranging from high frequency trading to behavioural herding at longer time scales. These mechanisms operate on different time scales, and have different magnitudes. Thus, the appropriate events must be defined to capture (and hopefully isolate) the dynamics of the mechanism of interest. Below, we present a non-exhaustive review of modern financial applications of Hawkes and ACD models.

As discussed in the introduction, high-frequency applications of Hawkes and ACD models are by far dominant in modern financial econometrics (see also~\cite{Hautsch2012}). In the context of the description of the order-book formation process, events can be naturally defined as a sequence of individual transactions~\cite{Engle1998,Engle2000,Hewlett2006} or quotes~\cite{Engle1997}, or more detailed as a set of mutually-exciting processes of submission and cancellation of limit orders and submission of market orders~\cite{Large2007,Abergel2010,Toke2011,JedidiAbergel2013}. On the aggregate level, the last transaction price change can serve as a proxy for cross-excitation between different markets~\cite{Muzy2011hawkes,Muzy2013hawkes}. Following the modern literature on price impacts (see~\cite{BouchaudLillo2009} and references therein), \cite{FilimonovSornette2012_Reflexivity} and \cite{FilimonovUNCTAD_JIMF2013} suggested
mid-quote price as a better proxy for price movements and mid-quote price changes were used for 
the estimation of the endogeneity of the price dynamics. In~\cite{Bowsher2007} and~\cite{Bacry2013impact},
the co-excitation between market orders and mid-price changes was used to model market impact.

However, applications of self-excited point processes are not limited to microstructure events (that can be defined only using complete order flow or level-1 tick data). In case of regularly-spaced discrete time time series (such as minutely, hourly or daily price dynamics), events can be defined as some kind of ``extremes'' in the dynamics. The most standard way (see for instance~\cite{McNeil2005,Embrechts2011} with respect to applications to daily data) defines events using the ``peak-over-threshold'' concept: for a given dynamics of financial returns, one selects those returns that fall outside a selected quantile range (for example 10\%--90\%). The resulting irregularly-spaced point process can then be calibrated using the Hawkes or ACD model. A more accurate approach should account for potential changes of regime and volatility clustering, and thus should use local extreme detection methods, such as the realized bi-power variation~\cite{Shephard2004} (\cite{BormettiLillo2013} apply this method to model co-jumps in time-series of 1-minute returns). 

Another interesting, but not yet explored application of point process models, involves detecting regime (or trend) changes in price dynamics and defining a point process using turning points. The simplest way is to define a local minima and maxima at a fixed time-scale in the discrete time series, and use these extrema to construct a point process. More accurate trend detection would involve local volatility estimation, such as method of drawup and drawdown detection (consecutive positive or negative price changes) discussed in~\cite{JohansenSornette2001JofRisk}. However, one needs to be warned that: (i) most trend detection methods are not causal and require information about the future price dynamics, thus they are not well-suited for forecasting purposes; and (ii) all these methods are based on conditional statistics that should be treated carefully
in order to avoid spurious phenomena even in featureless processes~\cite{FilimonovSornette2012_SpuriousTrendSwitching}. A general recommendation is to always consider 
one or several well-known processes (such as the uncorrelated random walk) and apply first the new
method to these known processes to check if the event defining procedure might not introduce 
some spurious endogeneity.
 
Finally, in modeling both micro- and macro-structure of financial time series, the magnitude of events (size of orders, size and sign of price changes or jumps) can be relevant.  In this case, a marked Hawkes model may be considered in which the size of the event determines its expected number of offspring, such as in the ETAS model
for earthquakes for which the marks are the earthquake magnitudes~\cite{VereJonesOzaki1982,VereJones1970,Ogata1988}.

%============================================================================== 
\section{Finite sample bias of the Hawkes maximum likelihood estimator}\label{apx:bias}

In order to optimize the calibration of the Hawkes model on the ACD($1$,$1$),
we study the finite sample bias and efficiency of the Hawkes maximum likelihood estimator. 
For this, we have simulated realizations of the Hawkes process with a modified thinning procedure~\cite{Lewis1979,Ogata1981} implemented in the same ``PtProcess'' package~\cite{Harte2010_PtProcess}, and afterwards we have calibrated the Hawkes model on this synthetic data. It should be noted that simulation (and fitting~\cite{Engle1998}) of the ACD model is computationally easier than for the Hawkes model.
Indeed, simulation of the Hawkes process with the thinning algorithm has complexity of $\mathrm{O}(N^2)$ (with possibility to reduce to $\mathrm{O}(N\log N)$ \cite{Moller2005,Moller2006}), compared
with complexity of only $\mathrm{O}(N)$ for the ACD($1$,$1$) model.

\bigskip

[Insert Figure~\ref{fig:Bias} here] 

\bigskip

We swept the parameter $\eta$ in the range $[0,1]$, fixing other parameters to $\mu=1$ and $\tau=1$. We generated 100 realizations of the Hawkes process of size 3500 each. To reduce the edge effects of the thinning algorithm, we discarded the first 500 points of each realization and afterwards calibrated the parameters of the Hawkes model on these realizations of length 3000. Figure~\ref{fig:Bias} illustrates the bias and efficiency of the maximum likelihood estimator in our framework. The definition of the Hawkes model~(\ref{eq:hawkes_discr}) requires the kernel $h(t)$ to be always positive. This implies $\eta\ge0$, so the estimation of $\eta$ is expected to have positive bias for small values, as seen in figure~\ref{fig:Bias}. On the other hand, when $\eta$ approaches the critical value of 1 from below, the memory of the system increases dramatically and, for critical state of $\eta=1$, the memory becomes infinite. Thus, for a realization of limited length, the finite size will play a very important role and will result in a systematic negative bias for $\eta\lesssim1$. This 
reasoning is supported by the evidence presented in figure~\ref{fig:Bias}, where one observes large systematic bias for $\eta>0.9$. For values of the branching ratio not too close to $0$ or $1$, 
the bias is very small for almost all reasonable realization lengths (longer than 200
to 400 points). We also find that the bias for $\eta$ close to $1$ strongly depends on the 
realization length. Finally, figure~\ref{fig:Bias} illustrates
the high efficiency of the maximum likelihood estimator: for values of $\eta<0.9$, the 
estimation error $|\hat\eta-\eta|$ measured with the 90\% quantile ranges does not exceed $0.1$.

%===============================================================================
%===============================================================================
\clearpage
%\bibliographystyle{apa-good-doi}
%\bibliography{/Users/vladimir/Work/Papers2/allpapers}

%===============================================================================
%===============================================================================
\pagebreak

\setcounter{figure}{0} \renewcommand{\thefigure}{\arabic{figure}}
\setcounter{table}{0} \renewcommand{\thetable}{\arabic{table}}

	\begin{table}[!ht]
	\caption{Estimated parameters of the Hawkes model with exponential~\eqref{eq:exp} and power law~\eqref{eq:pow} kernels together with values of log-likelihood ($\log L_{exp}$ and $\log L_{pow}$) and Akaike information criterion ($AIC_{exp}$ and $AIC_{pow}$) for cases presented in figure~\ref{fig:ACDHawkesSeries}. Bold font identifies the lowest AIC value among the two models.}
	\begin{center}
	\makebox[\textwidth][c]{
	\footnotesize
	\begin{tabular}{ccccccccccc}
	\toprule
	& $\alpha$ & $\beta$ & 
	$\theta_{H,exp}=(\hat\mu,\hat n, \hat\tau)$ & 
	$\theta_{H,pow}=(\hat\mu,\hat n, \hat c, \hat\varphi)$
	& ~ & $\log L_{exp}$ & $\log L_{pow}$ & ~ & $AIC_{exp}$ & $AIC_{pow}$ \\
	\cline{2-11}
	A~ & 0.05 & 0.05 & (0.84, 0.07, 4.3) & (0.83, 0.10, 262.77, 73.13) & 
		& $-343.0$ & $-343.4$ & & \textbf{692.0} & 694.8 \\
	B~ & 0.38 & 0.13 & (0.24, 0.52, 5.6) & (0.21, 0.54, 501.14, 113.41) & 
		& $-490.4$ & $-491.4$ & & \textbf{986.8} & 990.8 \\
	C~ & 0.13 & 0.38 & (0.38, 0.22, 7.9) & (0.36, 0.23, 816.41, 105.17) & 
		& $-509.3$ & $-509.3$ & & \textbf{1023.6} & 1026.6 \\
	D~ & 0.45 & 0.45 & (0.02, 0.83, 28.4) & (0.01, 0.87, 203.91, 7.70) & 
		& $-900.0$ & $-901.5$ & & \textbf{1806.0} & 1811.0 \\
	\bottomrule
	\end{tabular}	
	}
	\end{center}
	\label{tb:exp_pow}
	\end{table}

\clearpage

\begin{figure}[h!]
  \centering
  \includegraphics[width=0.9\textwidth]{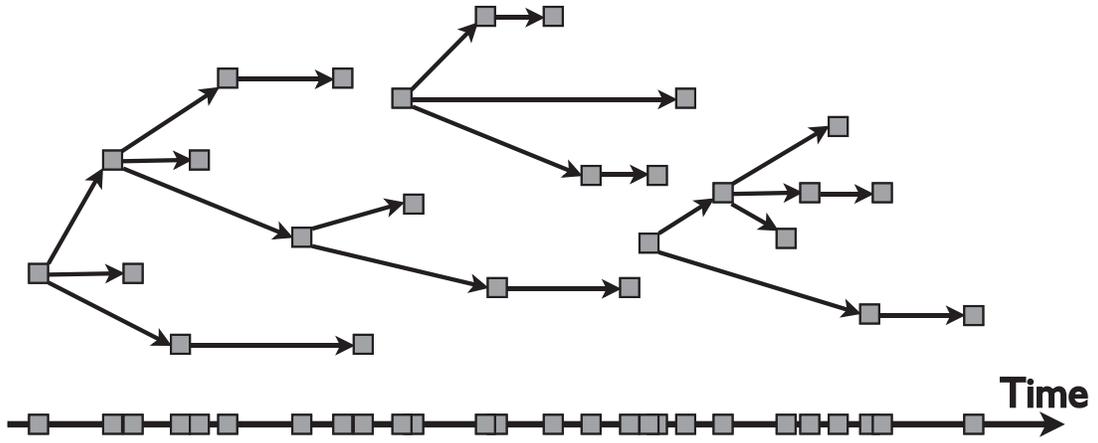}
  \caption{Illustration of the branching structure of the Hawkes process (top) and events on the time axis (bottom). This figure corresponds to a branching ratio $n=0.88$.} \label{fig:branching}
\end{figure}

\begin{figure}[hp!]
	\begin{center}
	\centerline{\includegraphics[width=0.9\linewidth]{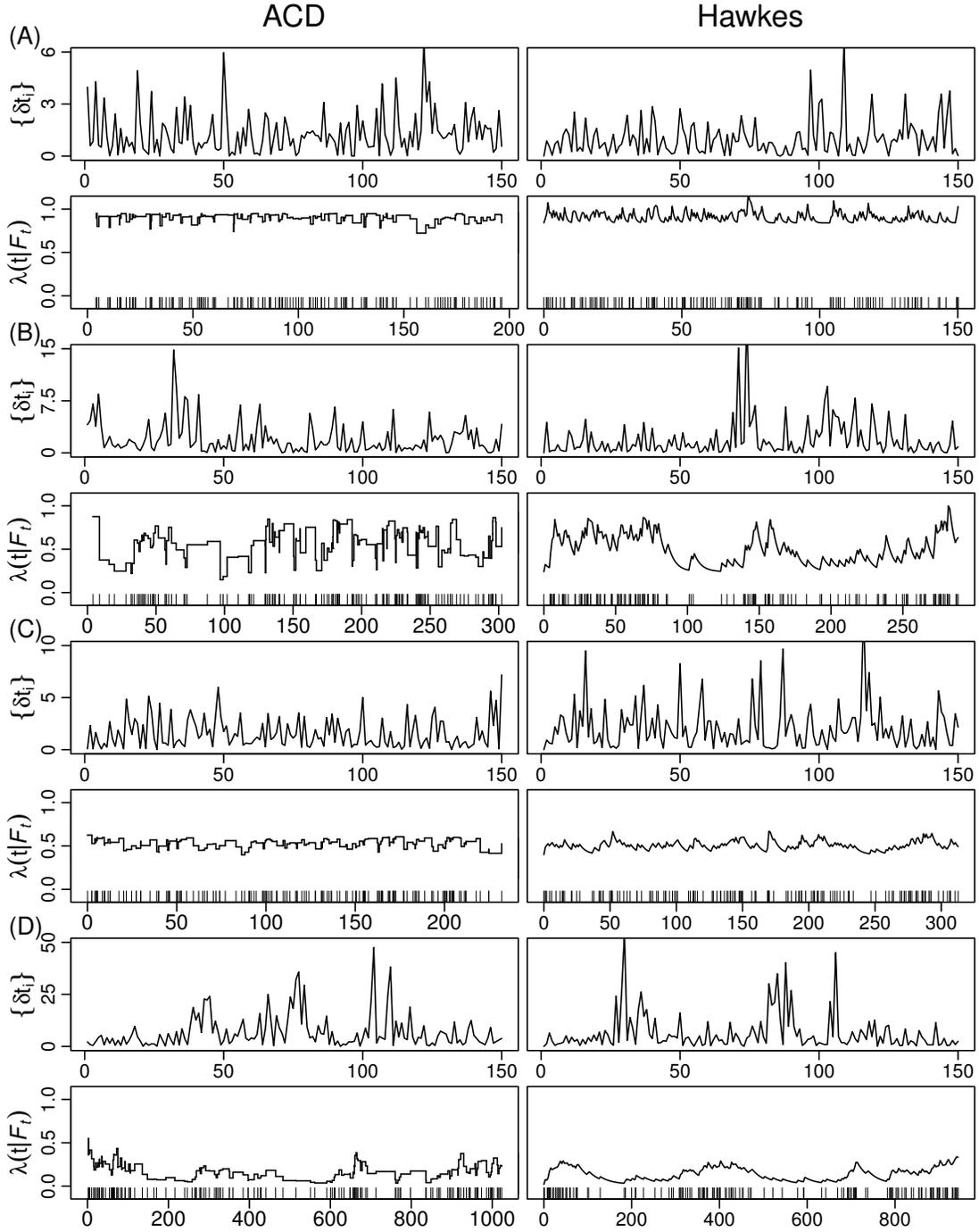}}
	\caption{Realizations of the durations and conditional intensities of the ACD($1$,$1$) process (left column), and Hawkes process (right column) simulated with parameters obtained by calibrating to the realization of the ACD process. Parameters of the ACD process $\theta_{ACD}=(\omega,\alpha, \beta)$ and estimated parameters of the Hawkes model $\hat\theta_{H}=(\hat\mu, \hat\eta, \hat\tau)$ are the following: 
	(A) $\theta_{ACD}=(1, 0.05, 0.05)$, $\hat\theta_H={(0.84, 0.07, 4.3)}$,
	(B) $\theta_{ACD}=(1, 0.38, 0.13)$, $\hat\theta_H={(0.24, 0.52, 5.6)}$,
	(C) $\theta_{ACD}=(1, 0.13, 0.38)$, $\hat\theta_H={(0.38, 0.22, 7.9)}$ and
	(D) $\theta_{ACD}=(1, 0.45, 0.45)$, $\hat\theta_H={(0.02, 0.83, 28.4)}$.}
	\label{fig:ACDHawkesSeries}
	\end{center}
	\end{figure}

\begin{figure}[t!]
	\begin{center}
	\centerline{\includegraphics[width=\linewidth]{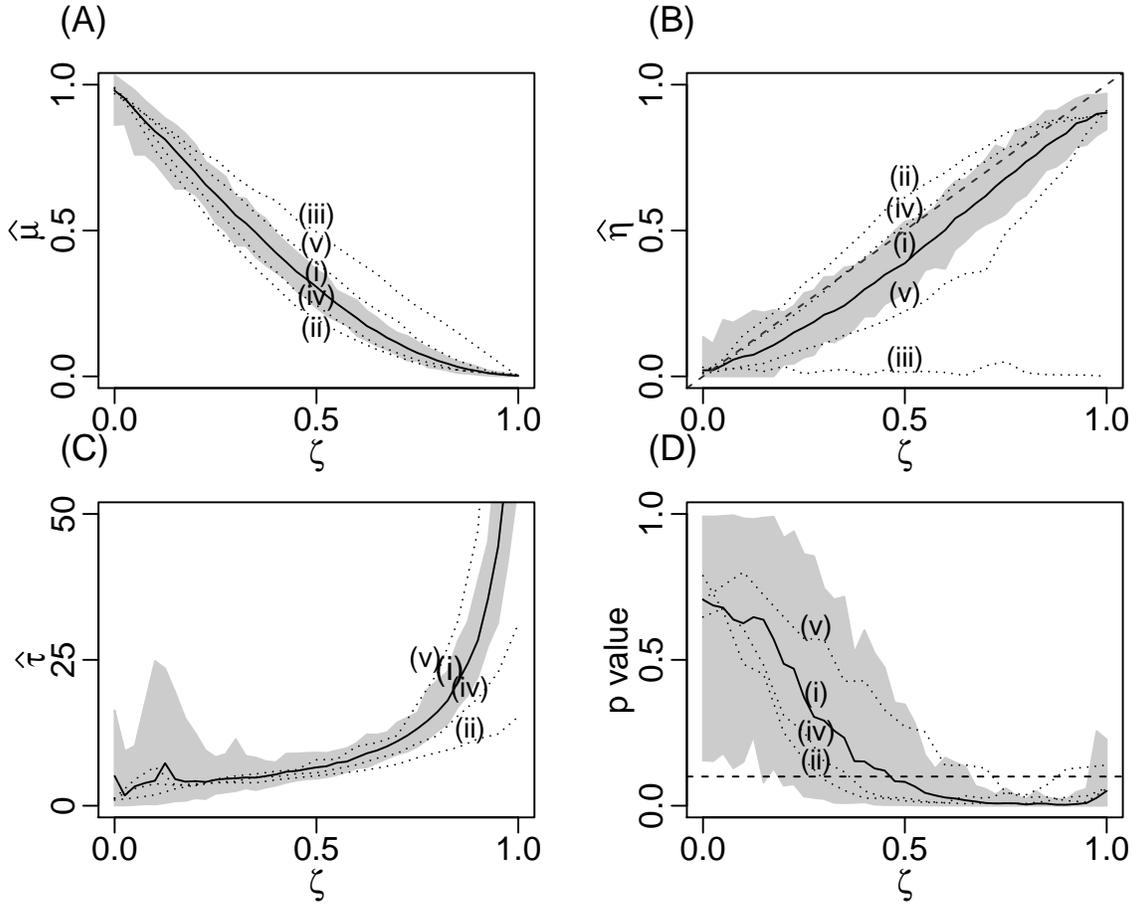}}
	\caption{Results of the calibration of the Hawkes model on the ACD($1$,$1$) realizations. Estimated (A) background intensity $\hat\mu$, (B) branching ratio $\hat\eta$, and (C) characteristic time of the kernel $\hat\tau$. Panel (D) shows the p-value from the goodness-of-fit test, where the dashed line 
indicates the the 10\% level.  (i) $\alpha=\beta~ (=\zeta/2)$, (ii) $\beta=0~ (\alpha=\zeta)$, (iii) $\alpha=0~ (\beta=\zeta)$, (iv) $\alpha=3\beta ~(=3\zeta/4)$ and (v) $\beta=3\alpha~ (=3\zeta/4)$.
The black line corresponds to the mean p-value 
for case (i) ($\alpha=\beta$), the shaded area to the 95\% quantile range for case (i), and the dotted lines depict mean p-values for cases (ii) $\beta=0$, (iii) $\alpha=0$, (iv) $\alpha=3\beta$ and (v) $\beta=3\alpha$.}
	\label{fig:HawkesFit}
	\end{center}
	\end{figure}

\begin{figure}[t!]
	\begin{center}
	\centerline{\includegraphics[width=0.7\textwidth]{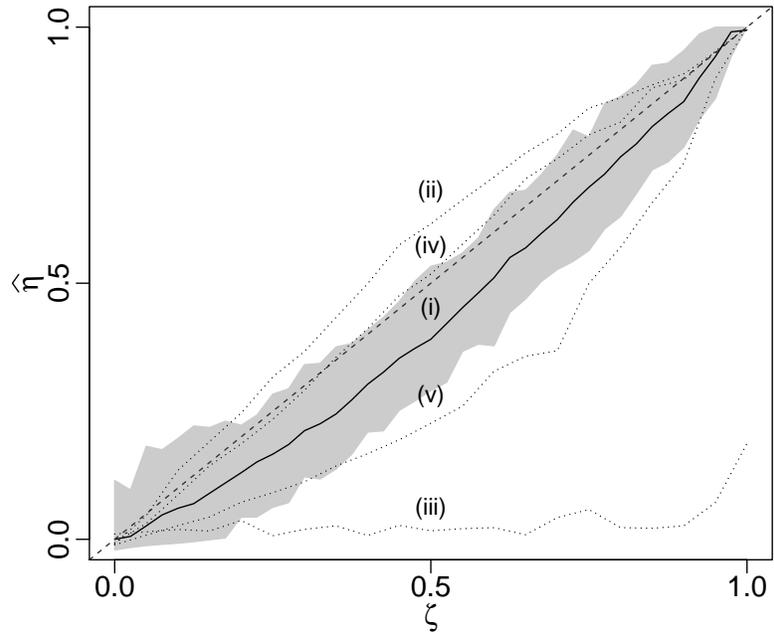}}
	\caption{The estimated branching ratio $\hat\eta$ of the Hawkes model estimated
	on ACD($1$,$1$) realizations with correction for the finite sample estimation bias determined in the~\ref{apx:bias}.}
	\label{fig:HawkesFitUnbias}
	\end{center}
	\end{figure}

\begin{figure}[t!]
	\begin{center}
	\centerline{\includegraphics[width=0.7\textwidth]{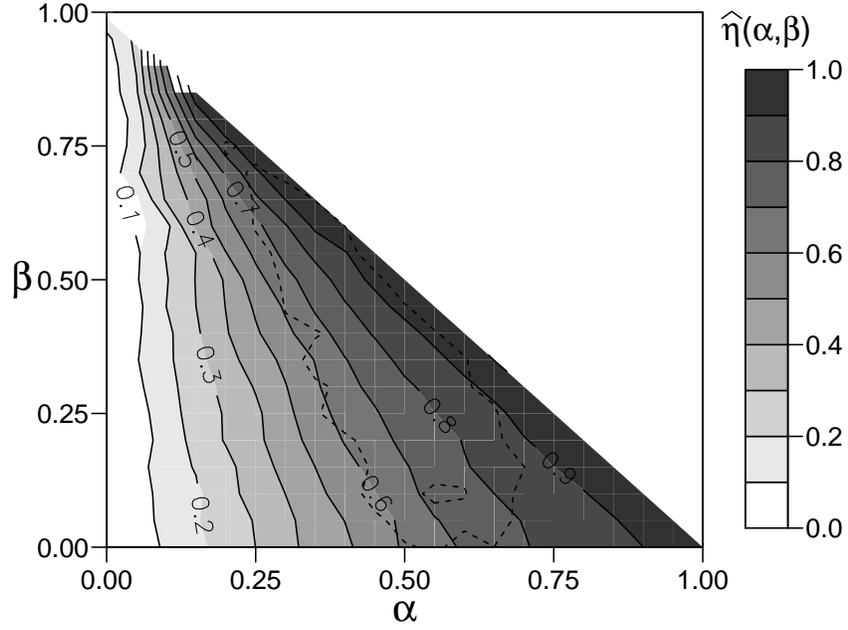}}
	\caption{Contour plot of the estimated branching ratio $\hat{\eta}(\alpha,\beta)$ of the Hawkes model 
	calibrated to ACD($1$,$1$) realizations for a grid of values $\alpha$ and $\beta$ with $\alpha+\beta\leq1$, corrected for the finite sample estimation bias determined in the~\ref{apx:bias}.  The dashed line 
	delineates the region where the goodness-of-fit tests rejects the null hypothesis (see text).}
	\label{fig:contour}
	\end{center}
	\end{figure}

	\begin{figure}[ht!]
	\begin{center}
	\centerline{\includegraphics[width=0.7\textwidth]{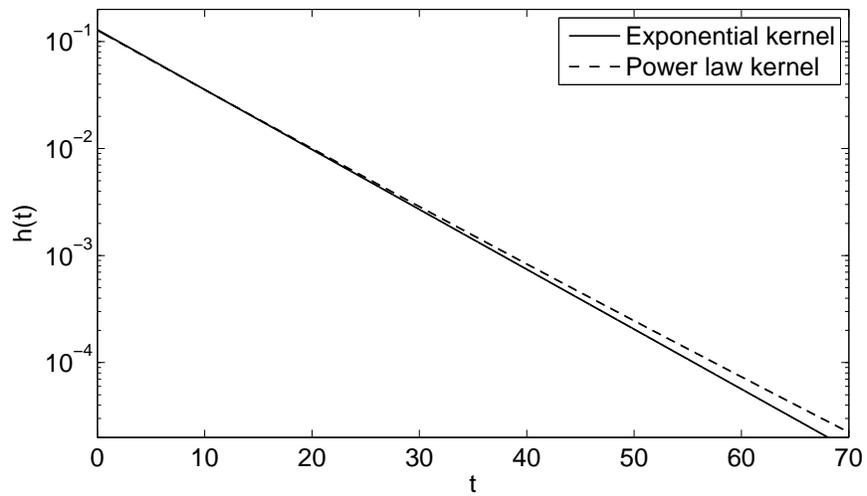}}
	\caption{Comparison of the exponential kernel with parameter $\tau=7.9$ (solid line) with the power law kernel with parameters $\varphi=105.17$ and $c=816.41$ (dashed line).}
	\label{fig:kernels}
	\end{center}
	\end{figure}

\begin{figure}[h]
	\begin{center}
	\centerline{\includegraphics[width=\linewidth]{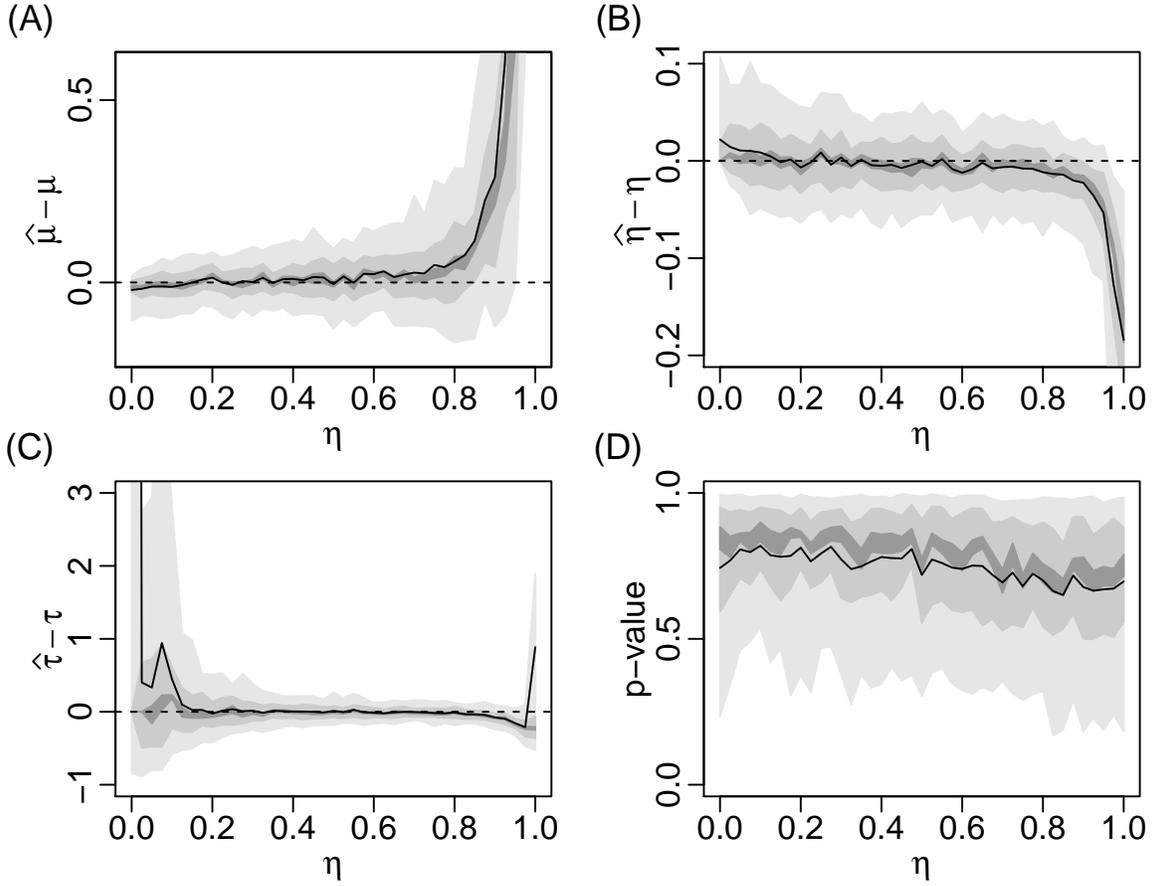}}
	\caption{ Illustrations of the finite sample bias and variance of the maximum likelihood estimator~\cite{Ozaki1979} of the parameters of the Hawkes process calibrated on time series generated
	by the Hawkes process itself (no model error).
	 Panel (A): difference between the estimates of the background intensity $\hat\mu$ and the
	 true value $\mu$ used for the generation of the time series; Panel (B): difference between the estimates of the branching ratio $\hat\eta$ and the true value $\eta$; Panel (C): difference between the estimates of the characteristic time of the kernel $\hat\tau$ and 
	 the true value $\tau$. Panel (D) 
	 shows the p-value of the Kolmogorov Smirnov test for standard uniformity of the transformed durations of the residual process \cite{Ogata1988}. In all panels, the black lines correspond to the mean, and the shaded areas to 90\%, 50\%, and 10\% quantile ranges. }
	\label{fig:Bias}
	\end{center}
	\end{figure}

\end{document}